\documentclass[12pt]{article}
\usepackage{graphicx}
\usepackage[T1]{fontenc}
\usepackage{floatrow}
\newcommand\pubnumber{NuPhys2015-Gollapinni}
\newcommand\pubdate{\today}

\def\ksu{Department of Physics\\
Kansas State University, Manhattan, Kansas, U.S.A.}
\def\support{\footnote{Work supported by U.S. DOE grant \textit{DE-SC0011840}.}} 

\def\Title#1{\begin{center} {\Large #1 } \end{center}}
\def\Author#1{\begin{center}{ \sc #1} \end{center}}
\def\Address#1{\begin{center}{ \it #1} \end{center}}

\newcommand\pubblock{\rightline{\begin{tabular}{l} \pubnumber\\
         \pubdate  \end{tabular}}}
\newenvironment{Abstract}{\begin{quotation}  }{\end{quotation}}
\newenvironment{Presented}{\begin{quotation} \begin{center} 
             PRESENTED AT\end{center}\bigskip 
      \begin{center}\begin{large}}{\end{large}\end{center} \end{quotation}}
\def\Acknowledgements{\bigskip  \bigskip \begin{center} \begin{large}
             \bf ACKNOWLEDGEMENTS \end{large}\end{center}}

\def\beq{\begin{equation}}
\def\eeq#1{\label{#1}\end{equation}}
\def\eeqn{\end{equation}}

\def\beqa{\begin{eqnarray}}
\def\eeqa#1{\label{#1}\end{eqnarray}}
\def\eeqan{\end{eqnarray}}

\let\bar=\overbar

\def\Dslash{\not{\hbox{\kern-4pt $D$}}}
\def\dslash{\not{\hbox{\kern-2pt $\del$}}}

\def\msb{{\bar{\ssstyle M \kern -1pt S}}}

\begin{document}
\begin{titlepage}
\pubblock

\vfill
\Title{Neutrino Cross Section Future}
\vfill
\Author{Sowjanya Gollapinni\support}
\Address{\ksu}
\vfill
\begin{Abstract}
The study of neutrino-nucleus interactions has recently received renewed attention due to their importance in interpreting the neutrino oscillation data. Over the past few years, there has been continuous disagreement between neutrino cross section data and predictions due to lack of accurate nuclear models suitable for modern experiments which use heavier nuclear targets. Also, the current short and long-baseline neutrino oscillation experiments focus in the few GeV region where 
several distinct neutrino processes come into play resulting in complex nuclear effects. Despite recent efforts, more experimental input is needed to improve nuclear models and reduce neutrino-interaction systematics which are currently dominating oscillation searches together with neutrino flux uncertainties. A number of new detector concepts with diverse neutrino beams and nuclear targets are currently being developed to provide necessary inputs required for next generation oscillation experiments. This paper summarizes these efforts along with a discussion of future prospects for precision cross section measurements. 
\end{Abstract}
\vfill
\begin{Presented}
NuPhys 2015, Prospects in Neutrino Physics\\
Barbican Center, London, UK, December 16--18, 2015
\end{Presented}
\vfill
\end{titlepage}
\def\thefootnote{\fnsymbol{footnote}}
\setcounter{footnote}{0}

\section{Cross sections for oscillations}
The global neutrino physics program is currently focused on studying the open questions about neutrino oscillations such as precision measurements of neutrino mixing parameters, $\delta_{CP}$ measurement, neutrino mass ordering and sterile neutrinos. Neutrino oscillations can be studied by observing the energy and flavor spectra of a beam of neutrinos (e.g., from an accelerator) at the beam source (usually with a near detector), before oscillations have started, and at the far detector, after oscillations have occurred. In addition to understanding the beam precisely, oscillation measurements also require a thorough understanding of neutrino-nucleus interactions to accurately reconstruct the incoming neutrino energy and compare the near and far fluxes. When neutrinos interact with the target material in a detector, they interact with nucleons that are bound within nuclei; the heavier the nuclei, the larger the impact of the nuclear environment. 
\begin{figure}[h]
\centering
\includegraphics[height=2.1in, width=2.4in]{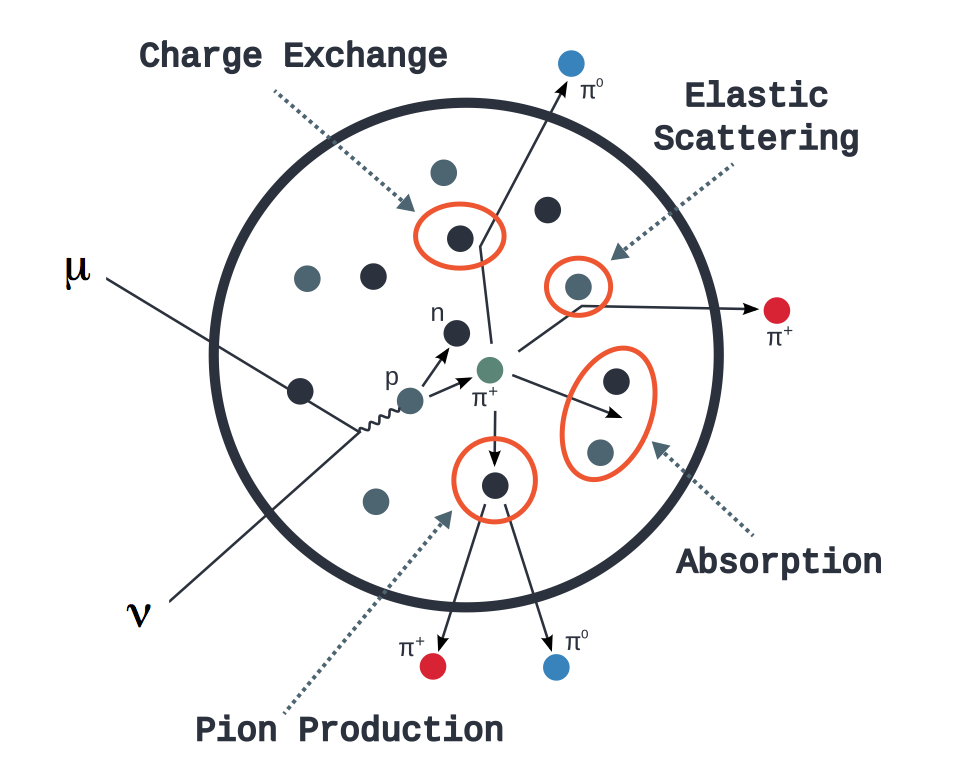}
\includegraphics[height=2.2in, width=2.5in]{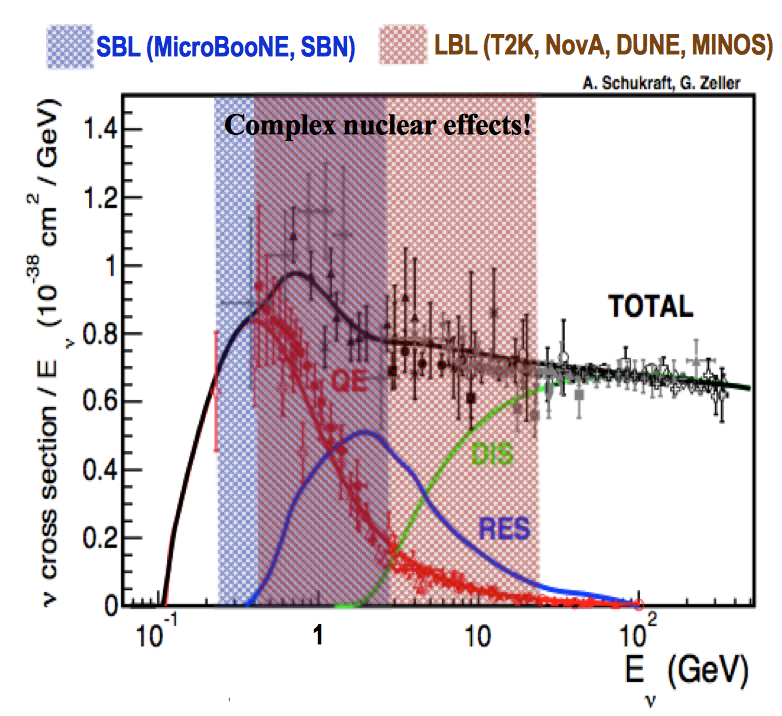}
\caption{(left) Illustration of how various processes get triggered when a neutrino interacts with a nucleus. (right) Neutrino energy landscape of current and future oscillation experiments.}
\label{fig:nuint}
\end{figure}
The universal scheme of using near detector (ND) data to constrain oscillation measurements in the far detector (FD) is not perfect due to oscillated flux and differences in E$_{\nu}$; usage of different detector technologies and nuclear targets can further complicate this scheme. Furthermore, the physics of neutrino oscillations depends on the initial neutrino state, and cross sections measured in the ND do not necessarily represent this due to flux uncertainties and detector effects. Also, to attain high statistics, modern neutrino experiments use heavier targets, where nuclear effects such as nucleon correlations and final state interactions (FSI) introduce significant complications and hadron kinematics come into play (see Fig.~\ref{fig:nuint}, left). For these reasons, experiments rely on nuclear models to convert the neutrino energy and flavor spectra detected at ND to initial interaction energy and spectra. Much of our understanding of neutrino scattering comes from data from light nuclei such as hydrogen and deuterium and so the existing models do not accurately represent the modern experimental data resulting in large systematics. Also, the statistical errors due to lack of data are also a big issue across the entire relevant energy range. More experimental input is needed to clarify and improve models of neutrino-nucleus interactions.

\section{Neutrino interaction systematics}
\begin{figure}[h]
\centering
\includegraphics[height=2.0in, width=2.3in]{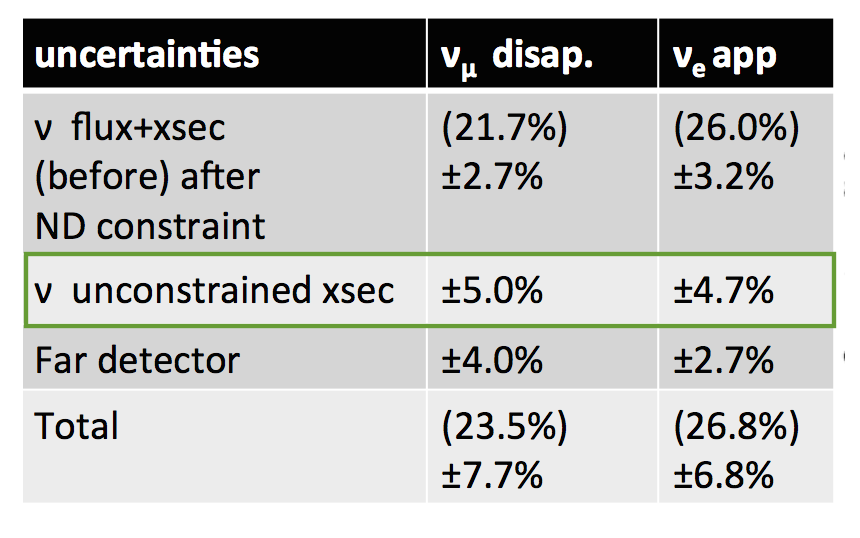}
\includegraphics[height=2.0in, width=2.5in]{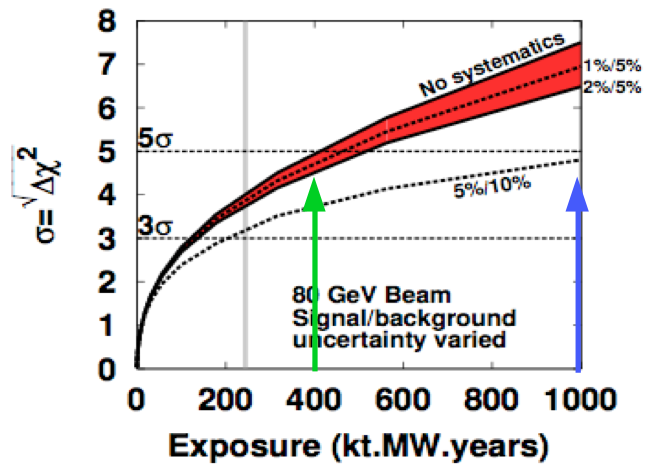}
\caption{(left) Table listing the uncertainty contributions from various sources for the T2K $\nu_{\mu}$ disappearance and $\nu_{e}$ appearance oscillation searches~\cite{T2Ksys}. (right) $\delta_{CP}$ sensitivity for 50\% of possible $\delta_{CP}$ values as a function of exposure (for a 12~MW and 34~kton detector) at the proposed DUNE experiment~\cite{mbass}. The green (blue) arrow represents 10 years (25 years) of running.}
\label{fig:t2k}
\end{figure}

Detection of $\nu_{\mu}\rightarrow\nu_{e}$ oscillations has been the recent focus of worldwide effort by short and long-baseline neutrino oscillation experiments as this channel depends on all the neutrino mixing parameters, including $\delta_{CP}$, and is also sensitive to non-standard oscillation physics.
Fig.~\ref{fig:nuint} (right) shows the energy landscape of current short and long-baseline neutrino oscillation experiments. All these experiments focus in the few GeV region where several neutrino processes contribute and nuclear effects are large. Neutrino interaction systematics are currently dominating oscillation searches at 		long-baseline experiments. As an example, 
Fig.~\ref{fig:t2k} (left) shows the cross section systematics at T2K for $\nu_{\mu}\rightarrow\nu_{e}$ oscillation searches. As seen from the table, the unconstrained neutrino cross section systematics at T2K is about 5\%~\cite{T2Ksys}. Similarly, the current cross section systematics for $\nu_{e}$ appearance searches at the NO$\nu$A experiment is around 7\%~\cite{novasys}. This current state of the art systematics of 5 to 7\% may not be good enough for future short and long-baseline experiments. 
\subsection{Impact of systematic uncertainties}
As an example, one can look at how the current state of the art systematics affect the $\delta_{CP}$ sensitivity at future long-baseline oscillation experiments such as DUNE. Fig.~\ref{fig:t2k} (right) shows the $\delta_{CP}$ sensitivity for 50\% $\delta_{CP}$ coverage as a function of the detector exposure for the proposed DUNE experiment~\cite{mbass}. For a 1.2~MW and 34~kton detector, the green arrow (blue arrow) represents a 10 year (25 year) run. The current state of the art systematics (5\%) results in 25 years of running to reach the 5$\sigma$ sensitivity whereas a systematics goal of 1\% makes it possible to achieve it in 10 years of running. Precision cross section measurements over the energy range valid for short and long-baseline neutrino experiments are vital for any oscillation measurement and as will be discussed in Section~\ref{sec:fut}, a number of current and future neutrino experiments are paving way for this.
\section{Roadway to precision cross sections}
In the current experimental scenario, there are many factors that affect precision neutrino cross section measurements, some of which are listed below:
\begin{itemize}
\item{Neutrino flux uncertainties are always a limit to precision cross section measurements. 
The current state of the art systematics for flux is $\sim$10\%~\cite{flux}. Hadron production data (e.g., from HARP) and dedicated $\nu$-electron scattering (such as from NO$\nu$A~\cite{novasys} and MINER$\nu$A~\cite{minervanue}) measurements can help reduce this further.} 
\item{Data on a broad range of nuclei (such as Fe, C, O, Pb, Ar) relevant for current and future oscillation experiments is needed to test nuclear models. More data on argon is especially needed in view of the number of liquid argon (LAr) based neutrino experiments that will become operational in the next few years. Also, cross section ratio measurements are desired as they provide more stringent tests of nuclear models due to cancellation of beam flux uncertainties.}
\item{Due to usage of heavier targets, modern experiments are moving towards event classification in terms of final state topology (e.g., CC 0-$\pi$, CC 1-$\pi$ etc.). This requires fine-grained detectors capable of detecting the hadronic side of the neutrino interaction. In the current scenario, proton identification threshold provides a good measure of detector capability. Liquid argon time projection chamber (LArTPC) neutrino experiments 
provide more data on argon and are also well-suited for precision $\nu$ cross section measurements due to their superior particle ID and excellent calorimetric reconstruction down to very low energy thresholds (the proton tracking threshold with the ArgoNeuT LArTPC is 21~MeV~\cite{argoneutNuclear}). LArTPCs are digitized bubble chambers that present neutrino interactions with unprecedented amount of detail. }
\end{itemize}
\noindent The next section will discuss future prospects for precision neutrino interaction measurements in the few GeV range at current and future neutrino experiments. 

\section{Future prospects}
\label{sec:fut}
\subsection{LAr-based experiments}

\subsubsection{ArgoNeuT}
\begin{figure}[h]
\centering
\includegraphics[height=1.4in, width=3in]{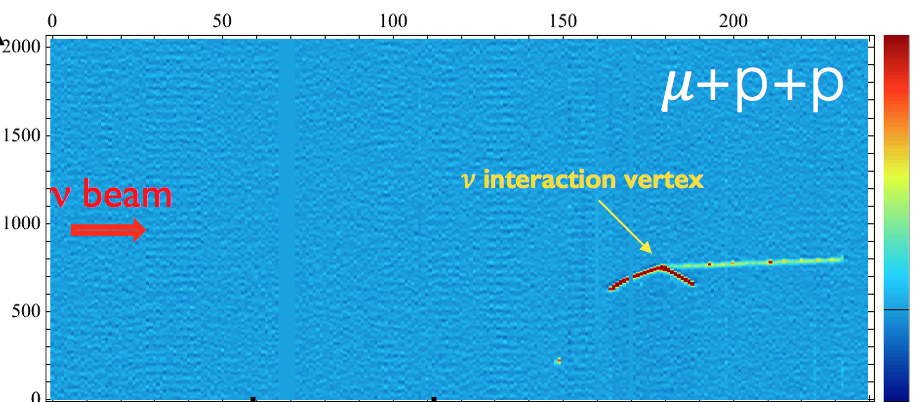}
\includegraphics[height=1.4in, width=2.1in]{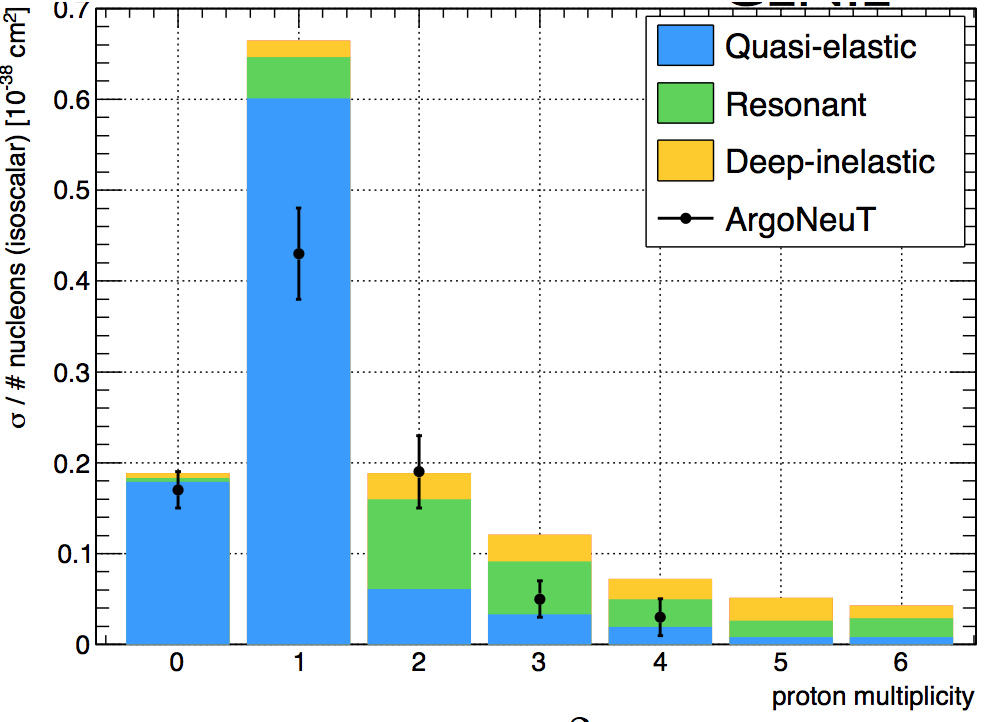}
\caption{(left) ArgoNeuT event display showing a multi-proton event. (right) CC 0-$\pi$ cross section on argon as a function of proton multiplicity from ArgoNeuT~\cite{argoneutnuint15}}
\label{fig:argo}
\end{figure}

The ArgoNeuT experiment~\cite{argonuet} on Fermilab's NuMI (Neutrinos at the Main Injector) beamline produced the first $\nu$-Ar cross section data in the medium energy region (3 to 10 GeV). Many impressive cross section measurements such as CC inclusive~\cite{argoneutCCincjosh, argonuetCCinc}, CC coherent pion production~\cite{argonuetCCcoh} and NC $\pi^{0}$~\cite{argonuetNCpi0} were produced in the last few years. ArgoNeuT also provided first direct investigations of nuclear effects by studying multi-proton events (see Fig.~\ref{fig:argo}, left) and events with back-to-back protons~\cite{argoneutNuclear}. Fig.~\ref{fig:argo} (right) shows recent results on CC 0-$\pi$ cross section as a function of proton multiplicity from ArgoNeuT~\cite{argoneutnuint15}. One can see a clear disagreement between data and MC (GENIE) in the plot arising from nuclear effects (for e.g., pion absorption) in argon. Also, given the small size of the detector, further analysis of exclusive states is statistically limited. Some of the current on-going analyses include $\nu_{e}$ CC and CC 1-$\pi$ cross sections. 

\subsubsection{MicroBooNE}
\begin{figure}[t]
\centering
\includegraphics[height=1.8in, width=2.6in]{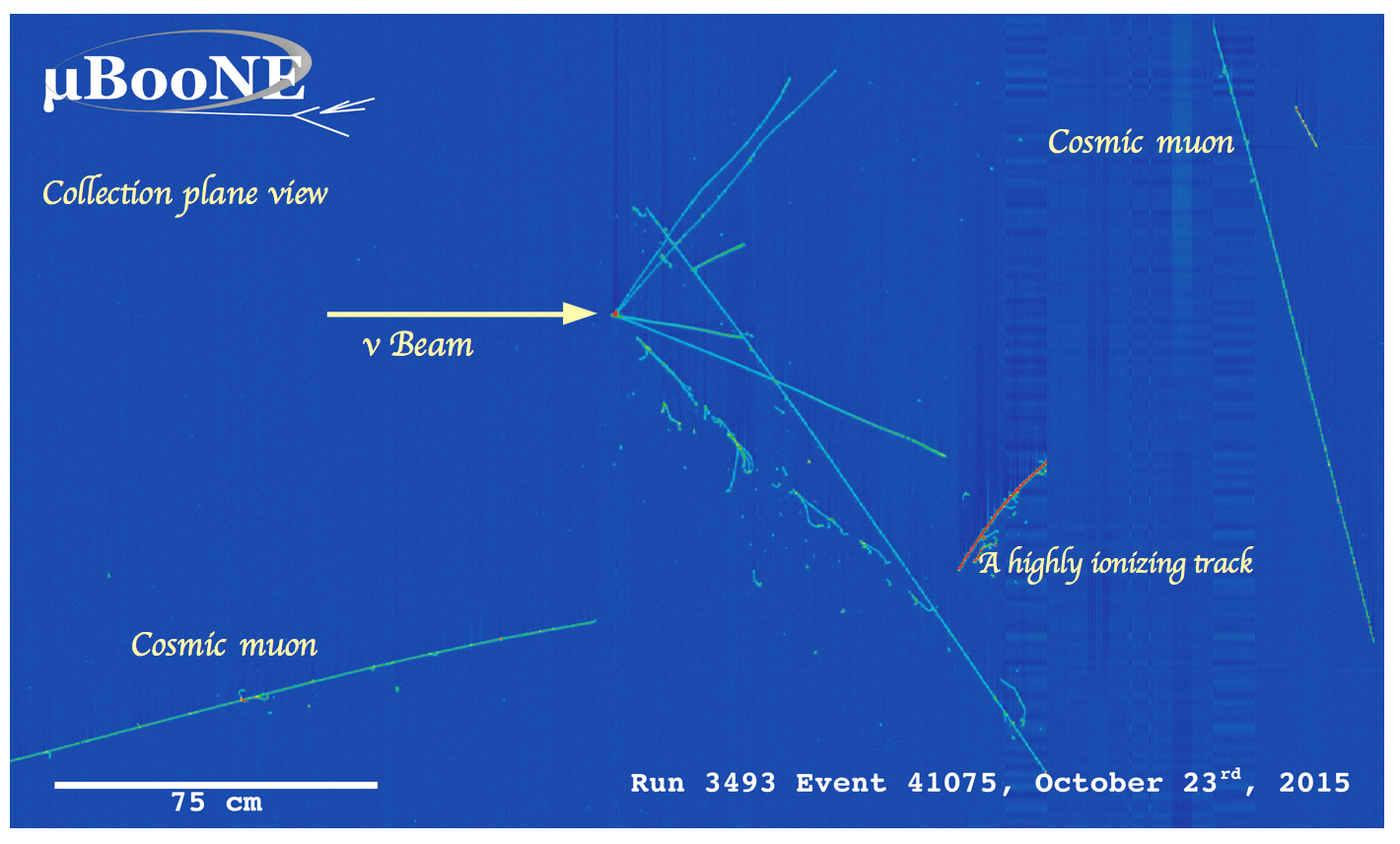}
\includegraphics[height=1.8in, width=2.6in]{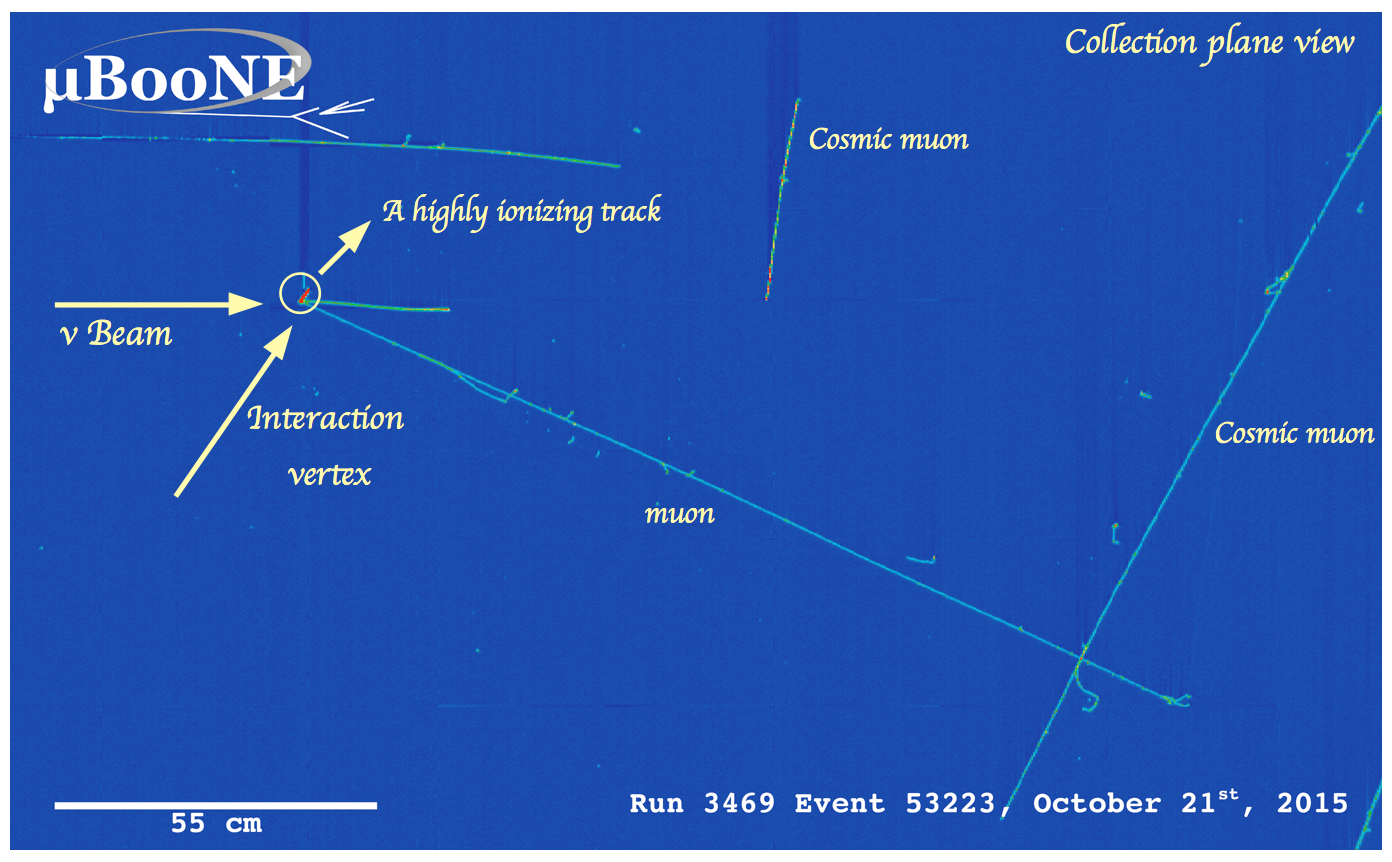}
\caption{Event displays showing neutrino interactions in MicroBooNE using recent BNB data.}
\label{fig:uBED}
\end{figure}

\begin{figure}[t]
\centering
\includegraphics[height=1.8in, width=1.9in]{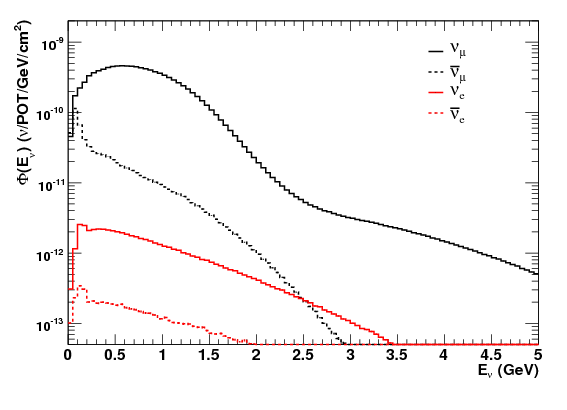}
\includegraphics[height=1.8in, width=1.9in]{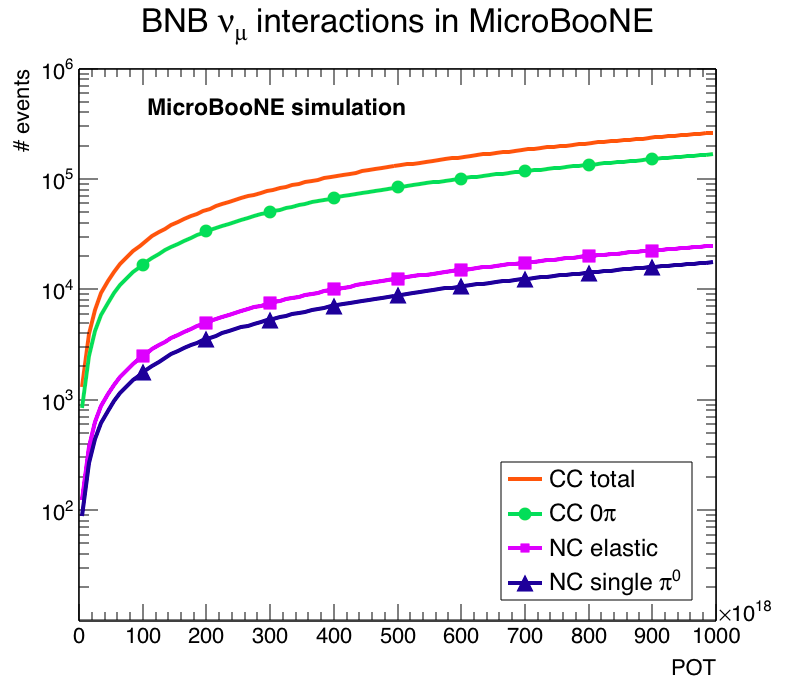}
\includegraphics[height=1.8in, width=1.9in]{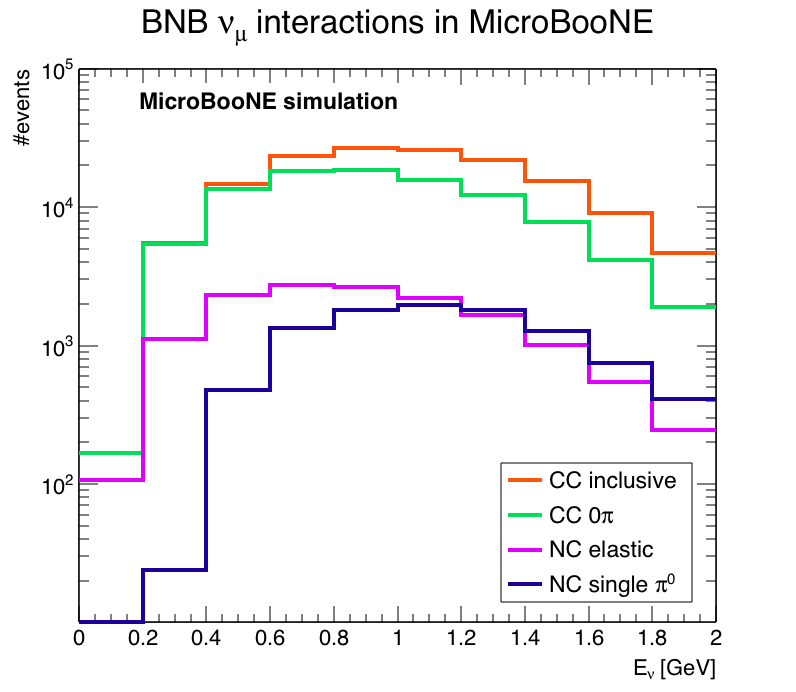}
\caption{(left) Expected BNB flux at MicroBooNE in the $\nu$ mode. (middle and right) Event rates in MicroBooNE for the BNB $\nu$ mode as a function of POT and E$_{\nu}$.}
\label{fig:uBStat}
\end{figure}

\begin{figure}[t]
\centering
\includegraphics[height=1.9in, width=2.8in]{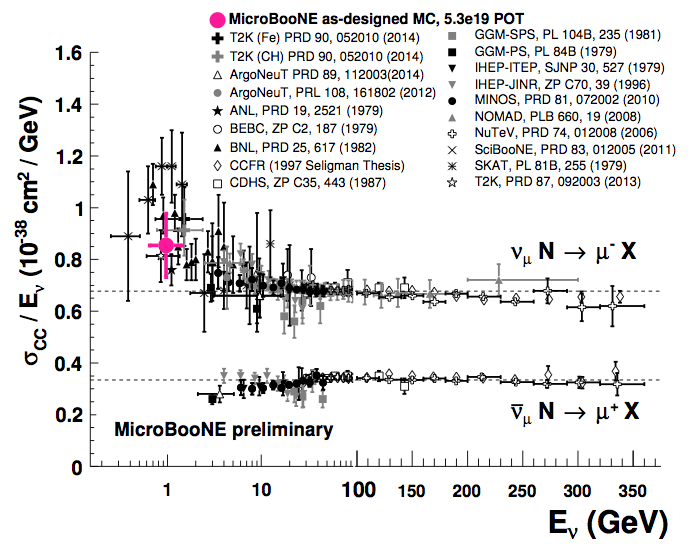}
\includegraphics[height=1.9in, width=2.1in]{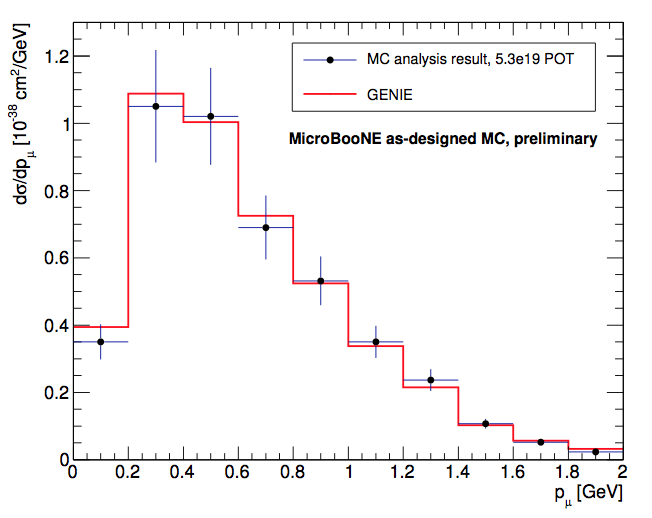}
\caption{(left) MicroBooNE flux-integrated $\nu_{\mu}$ CC cross section prediction~\cite{uBCC} derived from an as-designed detector MC compared to other data~\cite{PDG}.  (right) Differential cross section as a function of muon momentum~\cite{uBCC}.}
\label{fig:uBres}
\end{figure}

The next big avenue for neutrino cross section data on argon is the MicroBooNE experiment.
MicroBooNE is a 89-ton active volume LArTPC neutrino experiment built on the Fermilab Booster Neutrino Beam (BNB). MicroBooNE finished commissioning in Summer 2015 and has been collecting data with the BNB since October 2015 (see Fig.~\ref{fig:uBED}). Fig.~\ref{fig:uBStat} (left) shows the expected BNB flux at MicroBooNE as a function of neutrino energy. One of the main physics goals of MicroBooNE is to perform high-statistics precision measurements of $\nu$-Ar interactions in the 1~GeV range. MicroBooNE due to its large size will collect large statistics (see Fig.~\ref{fig:uBStat}, right) and will reach ArgoNeuT statistics in less than a month; most analyses are expected to be systematics limited in just a few months of data. For CC 0-$\pi$ statistics, MicroBooNE will be able to collect a total of 112,180 events in 3 years corresponding to a POT of 6.6E20. Similarly, in the case of 1$\mu$+2 proton exclusive channel, MicroBooNE will collect around 11,000 events in 3 years opening a great opportunity to study nuclear effects in argon. Fig.~\ref{fig:uBres} shows the sensitivity of MicroBooNE (as designed) for an early $\nu_{\mu}$ charged-current inclusive cross section analysis (both flux integrated and single differential cross sections) using 3 months of simulated BNB data~\cite{uBCC}. The MC results shown use contained muons and the event selection is entirely based on an automated event reconstruction. From this study, the flux systematics is expected to dominate the analysis in just 3 months of data collection. MicroBooNE has collected about 1.7E20 POT of data (as of 02/11/2016) and is re-estimating the CC inclusive sensitivity for the current state of the detector. A preliminary CC inclusive analysis result is expected in the next year from MicroBooNE. 

\subsubsection{Pion interaction cross sections from LArIAT}
\begin{figure}[h]
\centering
\includegraphics[height=1.5in, width=5.5in]{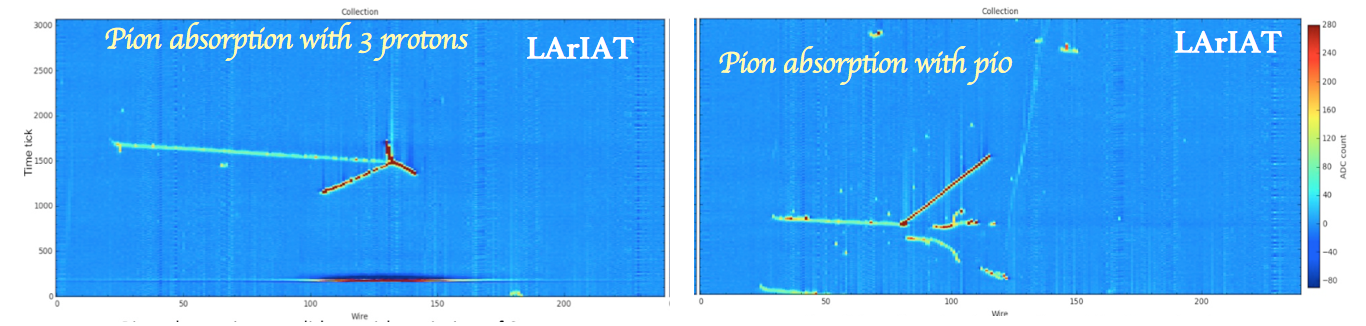}
\caption{Event displays showing $\pi$-Ar  interactions in LArIAT using recent data.}
\label{fig:lariat}
\end{figure}

The LArIAT (Liquid Argon In A Testbeam) experiment located at Fermilab 
uses a non-neutrino beam and is designed to explore the energy resolution and particle identification capabilities of LArTPCs in GeV-scale $\pi$-Ar interactions. The 
first LArIAT run took place from May to June 2015 and uses a tunable tertiary beamline produced from a high-energy pion beam (200~MeV to 1.5~GeV). The collected data will provide a great source for studying pion interaction cross sections (pion absorption, pion charge exchange, pion decay etc.) in argon (see Fig.~\ref{fig:lariat}) which haven't been studied in detail before. Pion absorption is a dominant effect for CC 0-$\pi$ cross sections in argon and understanding it is critical for future cross section measurements. Some of the current on-going analyses at LArIAT (with results aimed early next year) include total $\pi$ cross section, $\pi$ absorption and $\pi$ charge exchange. The next run, scheduled in Feb.~2016, aims to increase the statistics by six times using a high intensity beam and also tune the beam to increase kaon production by a few \% to study kaon interaction cross sections in argon. 

\begin{figure}[t!]
\CenterFloatBoxes
\begin{floatrow}
\ffigbox
  {\includegraphics[height=2.0in, width=2.5in]{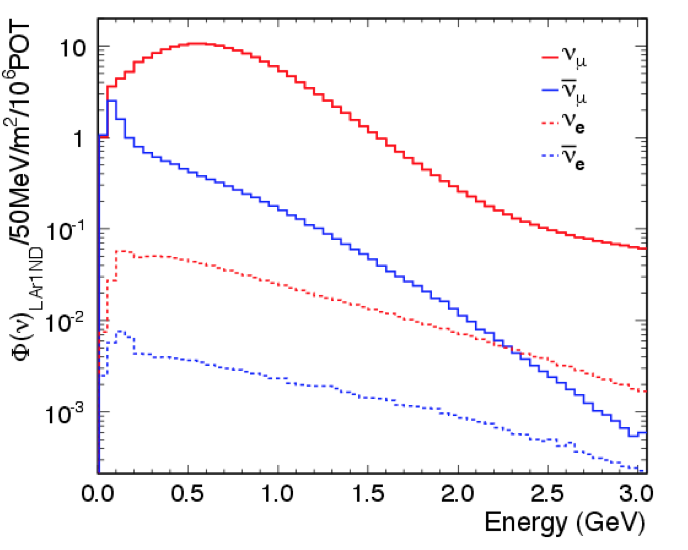}}
  {\caption{Expected BNB $\nu$ flux at SBND.}\label{fig:sbndflux}}
\killfloatstyle
\ttabbox
  { \begin{scriptsize}
   \begin{tabular}{|c|c|c|c|}
\hline
Channel & 1~month & 1~year &3~years \\ \hline
CC+Np&97,000&1,170,00&3,500,000\\
CC+0p&22,000&260,000&790,000\\
CC+1p&56,000&667,000&2,000,000\\
CC+2p&10,000&120,000&360,000\\
CC+($>$3p)&10,000&120,000&370,000\\
CC Coh. $\pi$&500&6,300&19,000\\
CC+1$\pi^{0}$&14,000&166,000&498,000\\
NC+1$\pi^{0}$&10,000&120,000&358,000\\
CC $\nu_{e}$&1,000&12,000&37,000\\
$\Lambda^{0}$ CC+NC&200&2,600&8,000\\
$\Sigma^{+}$ CC+NC&125&1,500&4,500\\
\hline
\end{tabular}
\end{scriptsize}
  }
  {\caption{Expected BNB event rates at SBND using GENIE neutrino generator.}\label{tab:sbndstats}}
\end{floatrow}
\end{figure}
\subsubsection{SBND}
The Short-Baseline Near Detector (SBND) is the near detector in the Fermilab Short-Baseline Neutrino (SBN) program. The design of SBND is almost final and the construction will start very soon with a goal to commission the experiment in 2018. Since SBND is located very close to the BNB beam (110~m from the BNB source), it will see about 30 times more $\nu$ flux compared to MicroBooNE and will be able to accumulate ArgoNeuT statistics in just two days. The expected impressive set of statistics (see Table~\ref{tab:sbndstats}) is complementary to MicroBooNE and will provide an excellent opportunity to perform high-statistics GeV-scale precision $\nu$-Ar cross section measurements including rare channels such as hyperon production in argon.

\begin{figure}[t!]
\centering
\includegraphics[height=1.9in, width=3.4in]{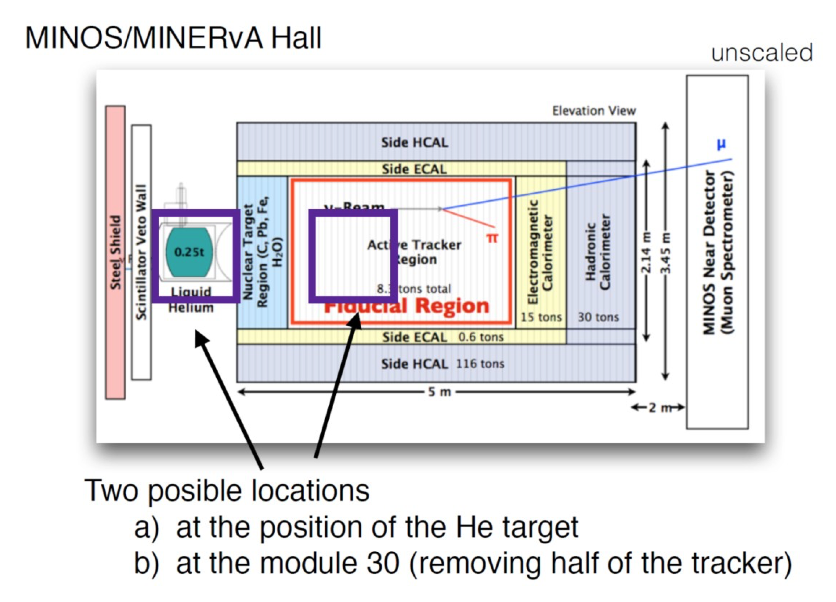}
\includegraphics[height=1.9in, width=2.5in]{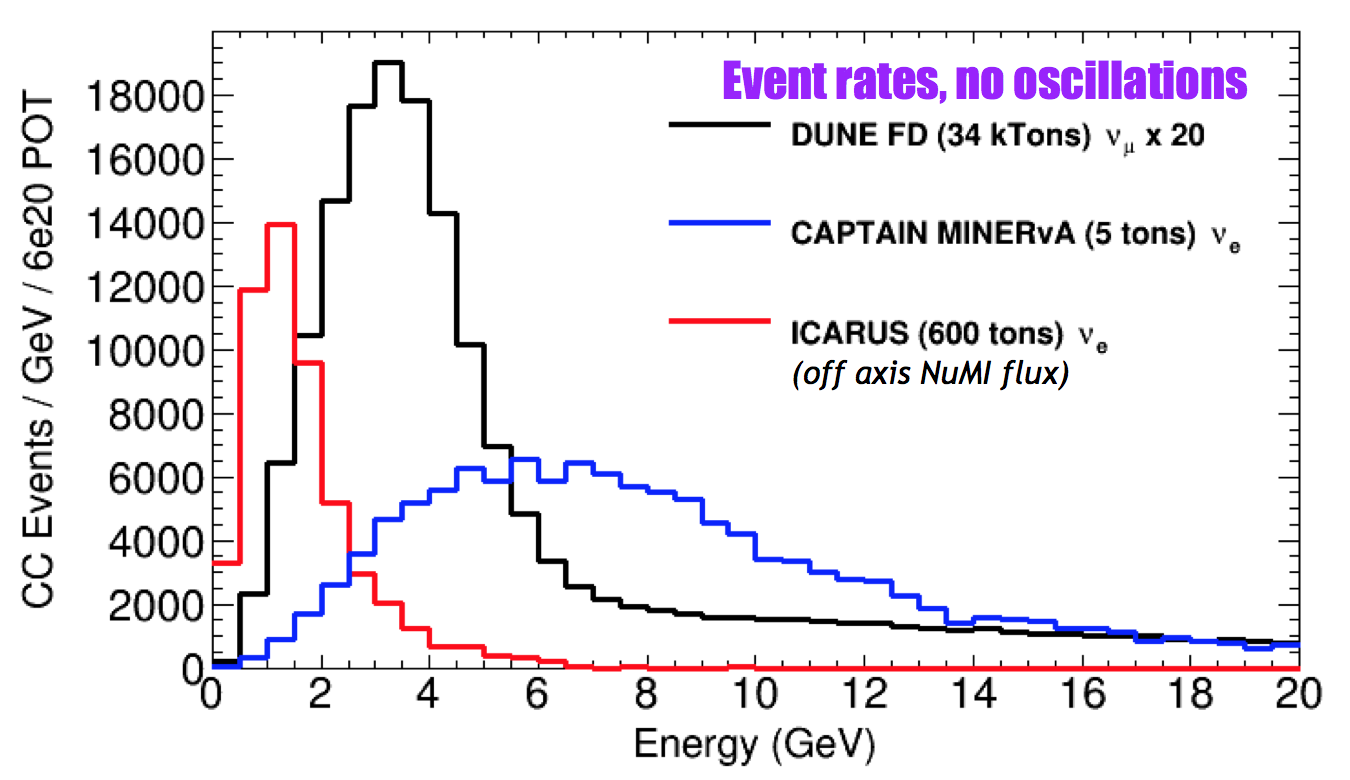}
\caption{(left) Possible locations for the CAPTAIN detector in the MINER$\nu$A experiment. (right) Expected CC $\nu_{e}$ event rates as a function of neutrino energy at CAPTAIN-MINER$\nu$A, compared to other experiments. The broad energy coverage of the CAPTAIN-MINER$\nu$A detector (blue line) complements the MicroBooNE and ICARUS programs~\cite{capmin}.}
\label{fig:capmin}
\end{figure}

\subsubsection{CAPTAIN-MINER$\nu$A}
The CAPTAIN-MINER$\nu$A program which received stage-1 approval in June 2015, extends the physics reach of MINER$\nu$A by placing 5~tons of active LAr target upstream of 6 or 9~ton scintillator target. It represents the second medium energy (3-10~GeV, NuMI beam) program in LAr after ArgoNeuT. Fig.~\ref{fig:capmin} (left) shows the two possible locations of the detector in the MINOS/MINER$\nu$A hall. The CAPTAIN-MINER$\nu$A detector will have 20 times more fiducial volume than ArgoNeuT resulting in more statistics and better containment (see Table~\ref{tab:capminstat}). CAPTAIN-MINER$\nu$A will measure cross sections across a broad range of energies/processes and will also be able to measure cross section ratios (e.g., Ar to CH) which will provide more stringent tests of nuclear models. With the prototype Mini-CAPTAIN successfully operating, the goal is to commission the CAPTAIN-MINER$\nu$A detector in 2018 with an initial plan to run for 2 years both in $\nu$ and anti-$\nu$ mode accumulating about 6.6E20 POT statistics.
\begin{table}[h]
\centering
 \begin{scriptsize}
\begin{tabular}{|c|c|c|}
\hline
Interaction&Events with reco. $\mu$ &Events with reco. $\mu$\\
channel&in MINOS/MINER$\nu$A&and charge in MINOS\\\hline
CCQE-like&460,000&390,000\\
CC 1$\pi^{+}$&980,000&480,000\\
CC 1$\pi^{0}$&780,000&300,000\\
\hline
\end{tabular}
\caption{Expected event rates at CAPTAIN-MINER$\nu$A in $\nu$ mode using GENIE for 6.6E20 POT~\cite{capmin}. A detector acceptance of 64\% is expected from MINOS (acting as downstream muon spectrometer) and MINER$\nu$A (acting as muon tracker) detectors.}
\label{tab:capminstat}
 \end{scriptsize}
\end{table}

\subsection{Non-LAr based experiments}
\subsection{MINER$\nu$A}
MINER$\nu$A~\cite{minerva} is a dedicated $\nu$ scattering experiment that collected low energy (peaking at 3~GeV) NuMI beam data from 2009 to 2012 and has many active analyses such as cross section ratio measurements~\cite{minerva2}, DIS on various nuclear targets~\cite{minervaDIS}, double differential $\nu$ and anti-$\nu$ QE scattering and kaon production cross sections.  MINER$\nu$A is currently collecting medium energy (peaking at 6~GeV) data with high statistics and will soon have interesting results in the new energy range. Some of the highest priority analyses with the medium energy data include neutrino DIS, CCQE, pion production and coherent pion production on scintillator, Fe, C \& Pb, $\nu$-e scattering on scintillator and nuclear structure functions.
\subsection{T2K}
\begin{figure}[h!]
\centering
\includegraphics[height=1.8in, width=2.2in]{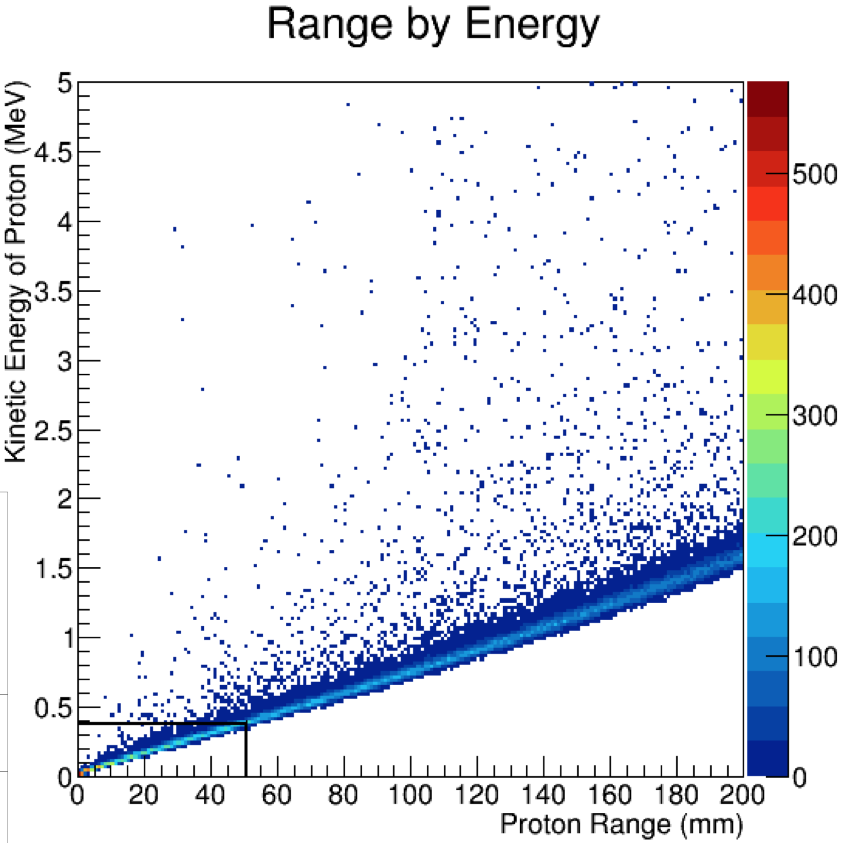}
\includegraphics[height=1.8in, width=3.5in]{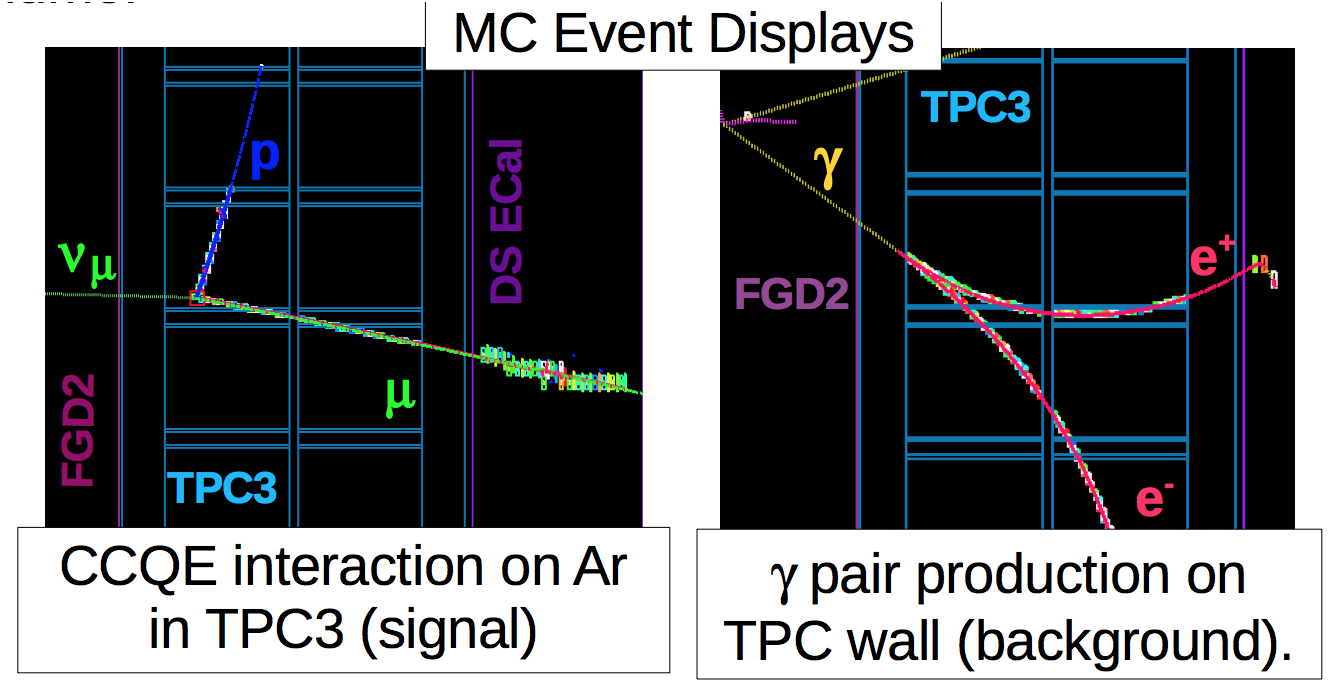}
\caption{(left) MC prediction of proton KE as a function of range in the ND280 TPCs. (right) MC event displays of $\nu$ interactions in the Ar gas in the ND280 TPCs.}
\label{fig:t2knd280}
\end{figure}

T2K recently presented a lot of impressive cross section results using the data from the two near detectors, INGRID \& ND280, at NuFact~2015~\cite{t2knufact} and NuInt~2015~\cite{t2knuint} conferences. There are analyses currently on-going on DIS cross section ratio measurements (e.g., Fe/CH) and to improve exclusive  
channel measurements which require better hadronic control as large FSI effects were observed. Also, there are on-going efforts to study FSI effects and measure CC inclusive flux integrated and differential cross sections in Ar gas of the off-axis ND280 near detector. The gaseous argon TPC allows interactions at a lower threshold than LAr. Fig.~\ref{fig:t2knd280} shows the kinetic energy of protons as a function of the proton range~\cite{t2knuintposter} and one can see that protons with energy as low as 0.5~MeV can be detected at ND280 (for comparison, ArgoNeuT was able to detect 21~MeV protons in LAr). Fig~\ref{fig:t2knd280} (right) shows some simulated event displays of interactions in ND280 gaseous argon TPC~\cite{t2knuintposter}.

\subsubsection{WAGASCI: T2K ND280 upgrade}
The non-canceling systematics due to different primary targets between the near (off-axis ND280 uses CH) and far (Super-K uses H2O) detectors is one of the main problems in T2K. To reduce $\nu$ interaction systematics for oscillation analyses, T2K will upgrade its off-axis ND280 detector with a plastic scintillator tracker with 3D grid like structure (see Fig.~\ref{fig:nds}, left). The goal of the WAGASCI project is to measure CC cross section ratio between H2O (signal) and CH (background) wit 4$\pi$ coverage and $<$~3\% systematic error. The current signal (H2O) to background (CH) ratio of 46:54 is expected to improve significantly (70:30) with the proposed upgrade. WAGASCI project was recently approved and is scheduled to commission in 2017.
\subsection{Other ND prospects}
\begin{figure}[h!]
\centering
\includegraphics[height=1.8in, width=2.3in]{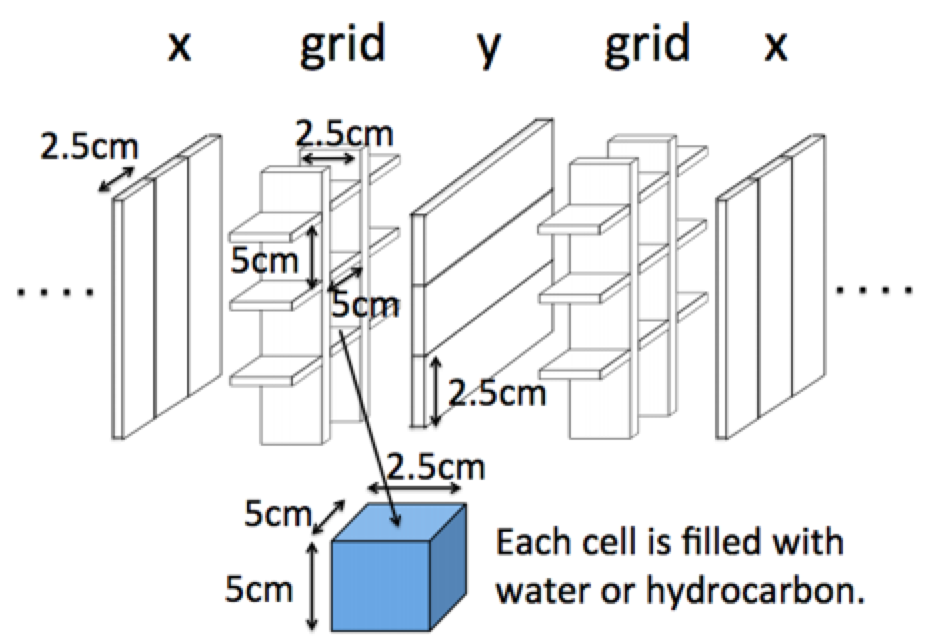}
\includegraphics[height=1.6in, width=1.6in]{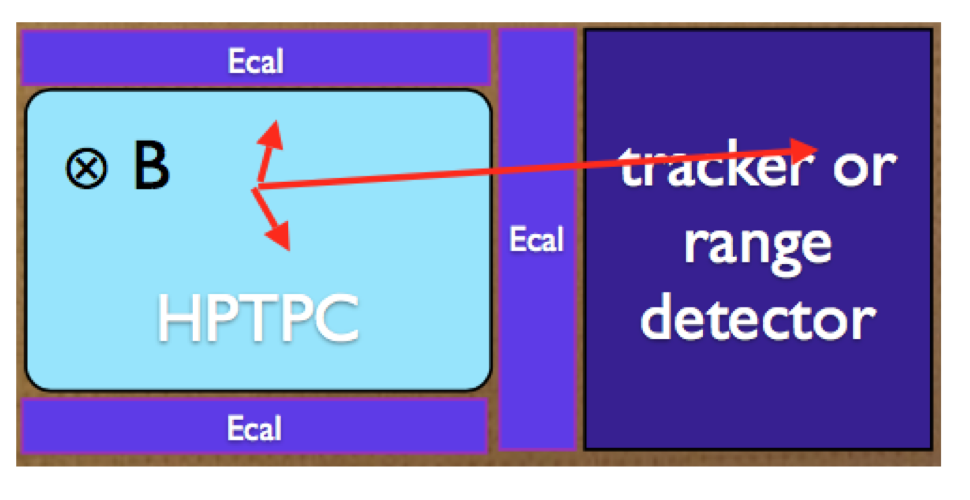}
\includegraphics[height=1.7in, width=2.0in]{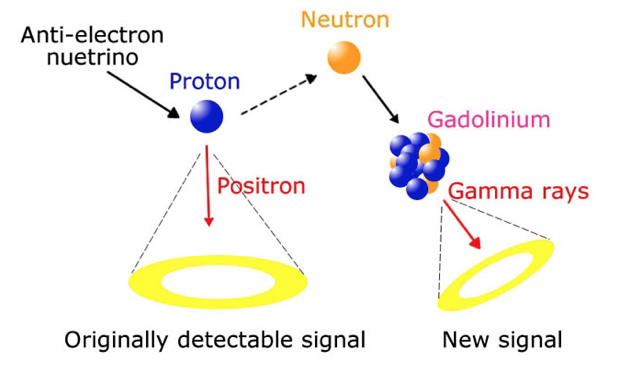}
\caption{(left) Proposed WAGASCI plastic scintillator tracker with 3D grid like structure. (middle) The principle of HPTPC. (right) Illustration of neutron capture in Gd doped targets.}
\label{fig:nds}
\end{figure}

Many near detector concepts are currently developed for DUNE, Hyper-K, and T2K. HPTPC is a 8~m$^{3}$ High Pressure gas TPC (see Fig~\ref{fig:nds}, middle) currently being developed for DUNE ND (also being considered for Hyper-K and T2K). The HPTPC possesses excellent PID capabilities and is sensitive to hadronic final states down to low thresholds. It can use He, Ne, Ar, and CF4 targets to study A-dependence of cross sections, multi-nucleon modeling and FSI. The current design is to surround the HPTPC by the ECAL for neutral particle containment. The ECAL could be made up of different target materials to allow for cross section ratio measurements.

Gadolinium (Gd) doping is a new technique currently being considered for near detectors to tag the presence of neutron (via neutron capture on Gd) in a final state to separate neutrino CCQE interactions from other processes ($\nu_{\mu}+n\rightarrow\mu^{-}+p$ versus $\overline{\nu}_{\mu}+p\rightarrow\mu^{+}+n$). This technique (see Fig.~\ref{fig:nds}, right) allows for high CCQE purity and improved anti-neutrino selection. Nuclear targets with 0.1\% Gd doping can capture neutrons with 90\% efficiency (Gd has 40,000~b capture cross section). The ANNIE (Accelerator Neutrino Nucleus Interaction Experiment) experiment on the Fermilab BNB is a Gd loaded water cherenkov (WC) detector which aims to measure neutron rates in GeV-scale $\nu$ interactions. Since this is in the same energy range as low energy atmospheric neutrinos, it will help reduce backgrounds from atmospheric neutrinos for next generation WC-based proton-decay search experiments such as Hyper-K. ANNIE is being considered as ND to Hyper-K program and recently received its phase-1 approval from Fermilab PAC.

\subsection{Intermediate detectors}
\begin{figure}[h!]
\centering
\includegraphics[height=2.0in, width=3.7in]{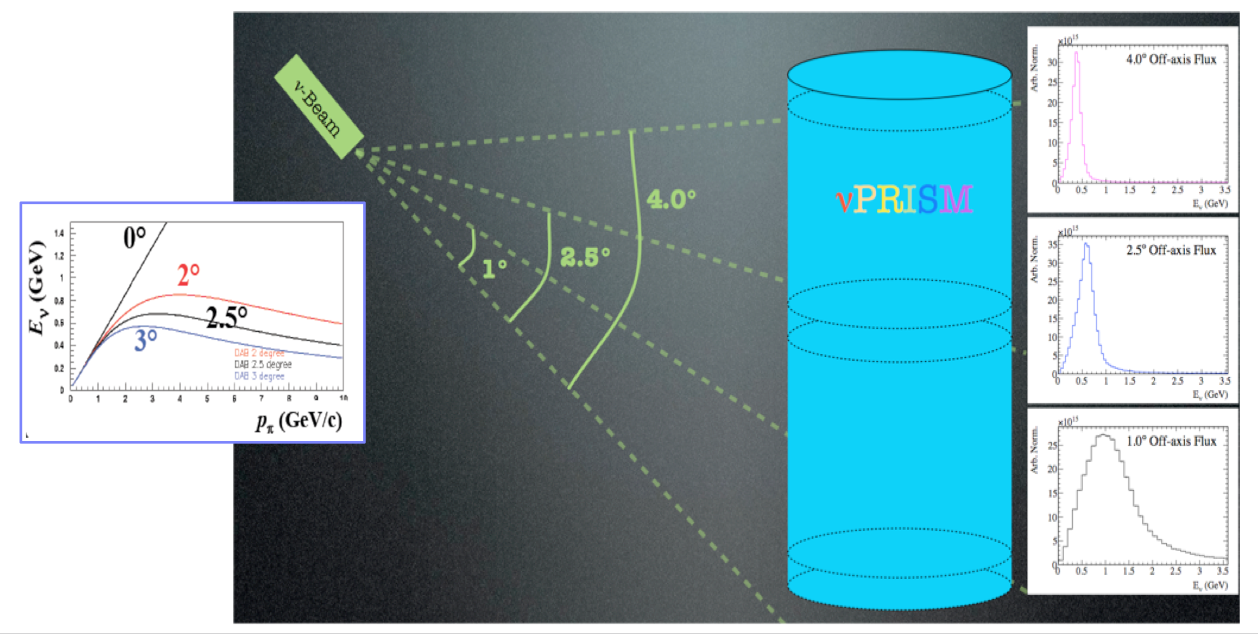}
\includegraphics[height=2.0in, width=2.2in]{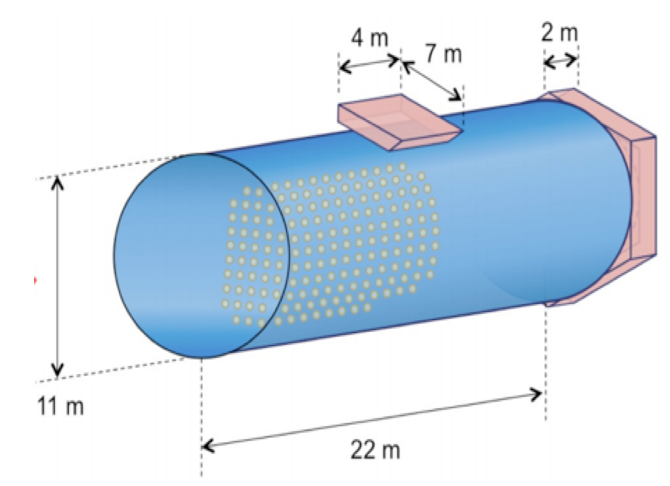}
\caption{(left) Generation of narrow band beams in NuPRISM. (right) Design of the TiTUS intermediate detector. }
\label{fig:nuprism}
\end{figure}

In addition to near detectors, intermediate detectors are also developed to reduce near to far extrapolation systematics. TITUS (see Fig.~\ref{fig:nuprism}, right) is a 2~kton Gd doped WC intermediate detector (2~km from the neutrino beam source) being proposed as ND for the Hyper-K program. NuPRISM is a kiloton-scale WC-based intermediate (1~km from the $\nu$ source) detector being considered for T2K. In accelerator neutrino experiments, flux is usually the dominant uncertainty in neutrino cross section measurements. This is due to the fact that accelerator neutrino beams are distributed on broad band spectra which results in E$_{\nu}$ to be reconstructed from observed final states. This makes cross section measurements very technology dependent. Monochromatic $\nu$ beams will allow one to probe the nucleus with a known energy and can greatly help reduce the systematics arising from beam.
NuPRISM is an off-axis spanning detector that takes advantage of 2-body decay kinematics from pions to create narrow band energy beam (1$\sigma$ width of 10\%). As shown in Fig.~\ref{fig:nuprism} (left), as off-axis angle increases, flux spectrum narrows producing narrow band energy beams. The JSNS$^{2}$ experiment will collect over 10$^{5}$ mono-energetic muon neutrinos (236~MeV) from charged kaon decay-at-rest for studying neutrino interactions and nuclear structure relevant for short and long-baseline experiments~\cite{josh}.

\section{Summary}
\begin{figure}[h!]
\centering
\includegraphics[height=1.8in, width=2.4in]{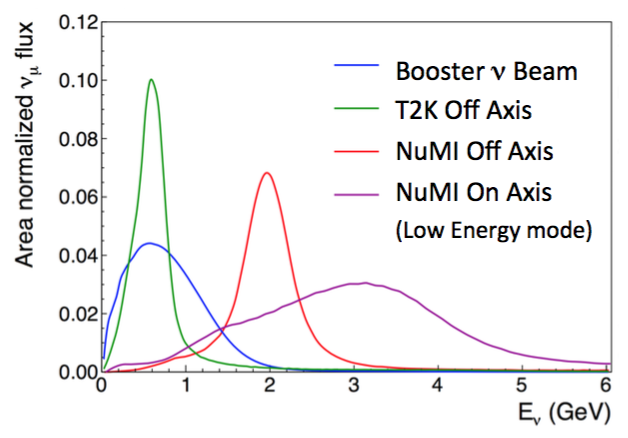}
\includegraphics[height=1.9in, width=2.4in]{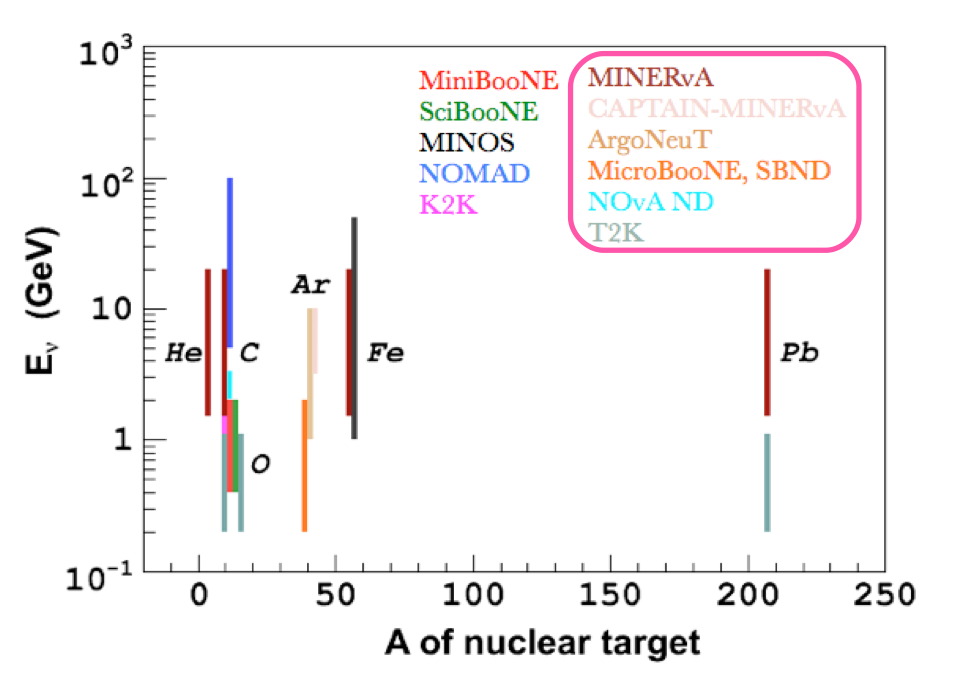}
\caption{(left) Energy spectra of various $\nu$ beams in the few GeV range currently used by oscillation experiments. (right) Modern neutrino cross section experiments~\cite{dave}.}
\label{fig:dave}
\end{figure}

A discussion of future prospects for cross section measurements is presented. Precision cross section measurements in the GeV-scale on nuclear targets relevant to current and future oscillation experiments are necessary to reduce neutrino interaction systematics and properly interpret oscillation results. With diverse beams (0.5~GeV to 6~GeV), different nuclear targets, and detector technologies, a rich cross section physics program (Fig.~\ref{fig:dave}) aimed towards achieving a complete picture necessary for future oscillation experiments is currently underway.
Experiments such as MINER$\nu$A, T2K and NO$\nu$A~\cite{nova} are continuing to produce interesting cross section results improving our understanding of the neutrino-nucleus interactions. The cross section measurements on argon in the GeV-scale with multi-ton detectors (MicroBooNE and LArIAT) will be available soon. In the near term (2018), SBND and CAPTAIN-MINER$\nu$A will become operational. Gadolinium loaded water cherenkov detectors (ANNIE, SK-Gd) are being developed to improve CCQE selection. Near (WAGASCI, HPTPC) and intermediate (TITUS, NuPRISM) detector concepts for T2K, DUNE and Hyper-K are being developed actively. The next few years will yield exciting cross section results greatly improving our knowledge on neutrino-nucleus interactions.

\Acknowledgements
The author thanks the organizers of NuPhys 2015 conference for the invitation to the conference. Special thanks goes to G.~Zeller, B.~Fleming, O.~Palamara and G.~Horton-Smith for their help in improving the conference presentation. The author would also like to thank J.~Raaf, F.~Cavanna, L.~Fields, D.~Harris, K.~Mahn, P.~Hamilton, P.~Shanahan and M.~Meuther for providing necessary materials on behalf of their respective scientific collaborations.

\end{document}